# Interface effects and dielectric mismatch in ultrathin silicon on insulator films

Andrea Pulici [a,b,*], Gabriele Seguini [a,*], Fabiana Taglietti [b], Roman Gumeniuk [c], Riccardo Chiarcos [d], Michele Laus [d], Johannes Heitmann [e], Marco Fanciulli [f], Michele Perego [a,*]

[a] *CNR-IMM, Unit of Agrate Brianza, Via C. Olivetti 2, 20864 Agrate Brianza, Italy*
[b] *Università degli Studi di Milano-Bicocca, Via Roberto Cozzi 55, 20125 Milano, Italy*
[c] *Institut für Experimentelle Physik, TU Bergakademie Freiberg, Germany*
[d] *Università del Piemonte Orientale ''A. Avogadro'', Viale T. Michel 11, 15121 Alessandria, Italy*
[e] *Institut für Angewandte Physik, TU Bergakademie Freiberg, Germany*
[f] *University of Torino, Department of Chemistry, Via P. Giuria 9, 10125 Torino, Italy*

[*]Corresponding Author:

andreapulici@cnr.it, gabriele.seguini@cnr.it, marco.fanciulli@unito.it, michele.perego@cnr.it

## Abstract

The role of interface states and dielectric mismatch is studied in ultrathin P-doped silicon-on-insulator (SOI) films with thickness of the device layer ($H_{SOI}$) varying from 30 to 8 nm and dopant concentration ($n_D$) ranging from $10^{18}$ to nearly $10^{20}$ cm$^{-3}$. P concentration is determined by Time-of-Flight Secondary Ion Mass Spectrometry (ToF-SIMS). Sample resistivity ($\rho$), carrier concentration ($n_e$), and mobility ($\mu_e$) are extracted by combining sheet resistance and Hall measurements in van der Pauw configuration. When $H_{SOI}$ = 30 nm, transport properties at room temperature are fully compatible with those of a similarly doped bulk Si. Progressive 2D confinement by reduction of $H_{SOI}$ below 30 nm results in a reduction of the carrier concentration and a concomitant degradation of $\mu_e$. These effects, which are steadily enhanced decreasing $n_D$, are attributed to non-passivated interface states at the SiO$_2$/Si interface and can be significantly mitigated by high temperature rapid thermal oxidation (RTO). The effectiveness of this approach was verified by electron-paramagnetic resonance (EPR) spectra and capacitance-voltage (CV) measurements, which allowed the assessment of the quality of the RTO-SiO$_2$/Si interface and the correlation with observed electrical properties. After effective interface engineering, low temperature electrical characterization revealed a significant increase in P ionization energy in samples with $H_{SOI} \leq 15$ nm, a result directly related to the dielectric mismatch.

# 1. INTRODUCTION

Silicon-based electronics rely on precise control of impurity atoms to tune the electrical and conducting properties of a semiconductor substrate. The incorporation of impurity atoms into the lattice of a semiconductor material, that is commonly indicated as doping, intentionally introduces donor or acceptor states within the bandgap of the semiconductor, shifting the Fermi energy level and modifying the effective carrier concentration in the material, ultimately enabling fundamental device functionalities such as *p-n* junctions, MOS capacitors, and photodetectors [1–3]. Over decades, advances in doping chemistry and processing have allowed researchers and engineers to tailor dopant profiles with fine depth and lateral resolution, both supporting the continuous scaling of microelectronic devices and promoting new architectures for the realization of CMOS devices, solar cells, and sensors [4]. Despite this maturity, achieving predictable activation, minimal dopant diffusion, and stable profiles, especially at ultrahigh or ultralow concentrations and in nanoscale geometries, remains a major challenge.

At the atomic level, donors (e.g., phosphorus, arsenic) and acceptors (e.g., boron) substitutionally incorporated in the Si lattice introduce discrete energy levels within the band gap very close to the conduction and valence band edges leading to temperature-dependent electrical ionization. Dopant-defect interactions and local chemistry play a major role in the effective activation of these dopants. In lightly doped Si, most dopants occupy substitutional sites and activate readily; however, at high concentrations, dopants interaction with the overlap of their electron wavefunctions induce the formation of energy bands that can alter ionization energy, mobility, and even the electronic structure of the semiconductors [5]. Above the solubility limit the impurity atoms form dopant clusters determining a significant deactivation of dopants themselves. Dopant diffusion, ion implantation, annealing, and defect engineering governs the final dopant distribution and activation, determining the electrical conductivity of the material. To achieve abrupt or highly localized concentration profiles, conventional approaches such as ion implantation followed by rapid thermal annealing, diffusion, and in-situ epitaxial incorporation are complemented by advanced methods including δ-doping, modulation doping, and non-equilibrium solid solubility [6–9].

The scenario is getting even more complex when considering doping of Si nanostructures because of the reduced dimensionality. Incorporation of dopant impurities in Si nanoclusters (NCs) with diameter below 10 nm is possible, even at concentrations well above the solubility limit [10–12]. However, the effective activation of these impurities and the availability of free charges in the Si NCs is questionable because of quantum confinement and surface related defects that may significantly alter the electronic behavior of these nanostructures. A comprehensive picture of the doping of these systems is still missing [13,14]. Si nanowires (NWs) present a distinct doping landscape due to their one-dimensional geometry and pronounced dielectric mismatch with the surrounding media that strongly modulates electrostatics and dopant activation and ionization[15]. Doping strategies include in situ axial or radial (core–shell) incorporation during bottom-up growth or post-growth methods such as ion implantation followed by tailored annealing, but small diameters (often tens of nanometers or less) hamper activation and promote surface-related dopant trapping, segregation to interfaces, or P complex formation with defects. Diffusion is effectively anisotropic and limited by the nanowire surface, enabling abrupt or graded profiles that can be engineered with radial dopant stacks or modulation doping along the wire. Dielectric mismatch fundamentally alters dopant ionization energies and carrier statistics through image-charge effects, increasing ionization barriers and modifying local band bending near the Si/dielectric interface [15,16]. The surrounding dielectric also governs gate coupling, screening, and Coulomb scattering, profoundly impacting mobility and threshold voltages in Si NW-based devices. Surface passivation and shell engineering (e.g., oxide or high-k dielectrics) are therefore crucial to suppress surface traps, tailor the local dielectric environment, and realize robust radial or axial doping schemes [17,18]. Actually, bulk-like resistivities can be retained to the atomic scale by fabricating "interface-free" dopant wires embedded in single-crystalline Si [19]. Collectively, these considerations connect atomic-scale dopant incorporation with macroscopic device performance in nanoscale transistors, sensors, and quantum-confined structures.

Silicon-on-insulator (SOI) substrates introduce a distinct context for silicon doping studies because the active device layer is electrically isolated from the bulk by a buried oxide (BOX). The thin Si device layer, with thickness from tens to a few hundred nanometers, enables strong electrostatic control and reduced parasitic capacitances but also makes dopant profiles highly sensitive to thermal processing and Si/SiO$_2$ interfacial effects. The BOX acts as a diffusion barrier that preserve abrupt vertical dopant gradients and suppress leakage into the handle wafer, yet it imposes a constrained thermal budget: excessive annealing can induce defect formation at the Si/BOX interface, cause film stress, or degrade interface quality [20,21]. Dopant ionization and mobility in ultrathin SOI with device layer thickness below 30 nm is severely influenced by the proximity and properties of the Si/BOX interface, as well as any residual strain, which can modify ionization energies and dopant solubility [22–24]. Ion implantation remains a quite common approach for impurity introduction into the Si device layer but requires careful dosage, proper energy selection, and specific beam geometry to minimize damage in the thin film and at the Si/BOX interface. Post-implantation annealing processes, such as rapid thermal processing, millisecond laser annealing, or solid-phase epitaxy, are often chosen to activate dopants while limiting diffusion across the device layer [25,26]. Additionally, the intrinsic back-gate capability and potential for strain engineering in SOI enable novel device architecture (e.g., fully depleted or stressed transistors) and back-gate-tuned dopant effects, highlighting the need to integrate dopant behavior with the unique electrostatics and thermal constraints of SOI [27,28]. Surprisingly, despite the broad technological interest for this semiconductor platform, very few studies systematically addressed the problem of doping of SOI substrates with ultrathin Si device layers [22,23].

In this work, we combine sheet resistance, Hall and capacitance-voltage (CV) measurements at room temperature to quantify how dopant dose, device layer thickness and processing conditions affect the device-relevant characteristics of the SOI substrate such as carrier concentration and mobility. Correlation of these data with dopant concentration profiles obtained by Time-of-Flight Secondary Ion Mass Spectrometry (ToF-SIMS) analysis is used to achieve information about effective dopant activation. Low temperature sheet resistance and Hall measurements are performed to determine ionization energy of the dopants as a function of their concentration and device layer thickness. By connecting atomistic behavior of dopants with macroscopic characteristics of the semiconductor material, this work aims to lay foundation of an empirical model for doping of ultrathin Si films in a wide range of concentrations and to provide robust fabrication strategies for next-generation Si technologies.

## 2. RESULTS AND DISCUSSION

**2.1 Sample Preparation and Compositional Analysis.** 1 × 1 cm² SOI dies with Si device layer having thickness ($H_{SOI}$) ranging from 30 to 8 nm were prepared and subsequently doped using polymers terminated with a P containing moiety [23,29]. Accurate control of the dose (cm$^{-2}$) of dopants injected into the Si device layer was achieved following two different doping approaches. The first approach relies on a methodology that was fully described in our previous publications: repeated grafting/ashing cycles of the polymer lead to a stepwise linear increase in the dose of P atoms grafted to the Si surface [29–31]. In the present work the number of grafting/ashing cycles varied between 1, 3, 5 and 10 to modify the amount of P in the dopant source. Upon deposition of a 10 nm thick $SiO_2$ capping layer, the samples underwent a single high temperature thermal treatment in an RTP system at $T$ = 1000 °C in $N_2$ atmosphere to promote the drive-in and redistribution of P atoms into the Si device layer. Accordingly, by properly adjusting the annealing time, it is possible to precisely control the effective amount of P atoms injected into the Si device layer and achieve a homogenous P concentration in the Si device layer. Thermal treatment at $T$ = 1000 °C for 100 s already demonstrated full activation of the dopants, uniform P concentration in the case of 30 nm thick SOI samples and optimal electrical properties [21].

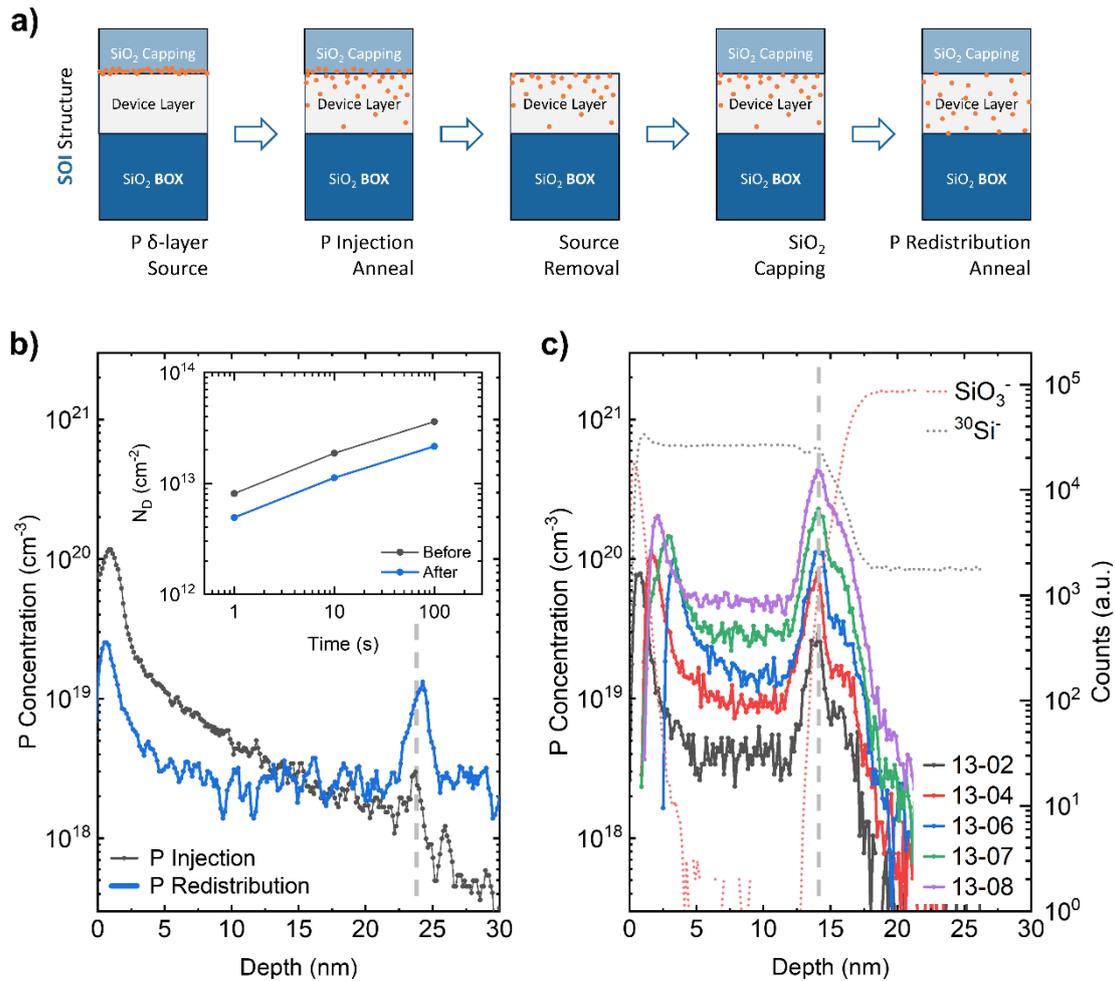

*Figure 1.* (a) Schematic representation of the "double annealing" doping approach. The first annealing selects the P dose injected into the device layer while the second uniformly redistributes and activates the dopants. (b) Calibrated ToF-SIMS P depth profiles of a SOI sample with $H_{SOI}$ ~ 23 nm obtained before (black) and after (blue) the redistribution anneal. In the inset, $N_D$ values before and after the redistribution annealing as a function of the duration of the injection treatment. (c) Calibrated ToF-SIMS P depth profiles of SOI samples with $H_{SOI} = 13 \pm 1$ nm doped following different sample preparation procedures. Detailed information about sample preparation is reported in Table S1. Grey dashed line indicates the position of the Si/BOX interface.

The second approach relies on an alternative protocol aiming to reduce P concentration into the Si device layer by decoupling the injection and redistribution processes, which are carried out in two distinct thermal treatments. A schematic representation of this so-called *double annealing* approach is reported in **Figure 1a**. A P δ-layer source is created at the surface of the Si device layer by means of a single grafting/ashing cycle. The first annealing process is performed at $T$ = 900 or

1000 °C in $N_2$ atmosphere. The duration of the annealing is selected to control the amount of P atoms injected into the Si device layer [21]. The calibrated P concentration profile obtained on a SOI sample with $H_{SOI}$ ~ 23 nm upon drive-in at $T$ = 1000 °C for 1 s is reported in **Figure 1b** (black line). The short duration of the injection treatment results in a P concentration gradient throughout the entire device layer. The observed concentration profile is correctly predicted by Fick's law of diffusion for thick SOI films[32]. Upon injection, a HF bath is performed to remove the 10 nm $SiO_2$ capping layer and the P δ-layer source. A new 10 nm thick $SiO_2$ capping layer is immediately re-deposited by e-beam evaporation to prevent out-diffusion of the P atoms. A second high temperature treatment is performed in an RTP system at $T$ = 1000 °C for 100 s in $N_2$ atmosphere to promote the redistribution of the dopants throughout the Si device layer. **Figure 1b** also reports a representative calibrated P profile obtained on the same SOI sample at the end of the double annealing process (blue line). Calibrated ToF-SIMS depth profiles demonstrate uniform dopant concentration throughout the entire Si device layer. Moreover, clear reduction of the P dose ($N_D$) confined in the device layer is observed after annealing. The inset of **Figure 1b** shows the total $N_D$ injected in the device layer, computed as the integral of the P concentration depth profile obtained by ToF-SIMS analysis. The duration of the first injection treatment at $T$ = 1000 °C was varied between 1, 10 and 100 s. $N_D$ values obtained before and after the second redistribution treatment clearly indicate a 40 % reduction of the dose of P dopants in the Si device layer upon second thermal treatment due to possible P segregation and out-diffusion through both the top oxide (TOX) and BOX interfaces [23].

Control of the P dose was demonstrated by the calibrated ToF-SIMS P depth profiles reported in **Figure 1c** even in the case of ultrathin SOI samples with $H_{SOI}$ = 13 ± 1 nm. Varying the processing conditions, the P atoms are redistributed uniformly throughout the Si device layer while the P concentration is varied in a wide range of dopant concentrations ($n_D$). A small increase in the P signal at the device layer/BOX interface was observed when increasing $n_D$, suggesting some P accumulation at the Si/BOX interface and diffusion in the BOX [23]. All the details regarding sample preparation of each P doped SOI sample are reported in **Table S1**.

**2.2 Electrical Transport Measurements.** SOI samples with 30 nm thick Si device layer and different P concentration were prepared by properly adjusting parameters during the doping process. The $n_D$ in the device layer of each sample was determined to be constant throughout the entire film thickness by ToF-SIMS analysis, with values varying between $10^{18}$ and $10^{19}$ cm$^{-3}$, depending on the processing conditions. The carrier concentration ($n_e$) in the samples is derived as the ratio between the total carrier dose ($N_e$), directly measured by Hall measurements in vdP configuration, and the $H_{SOI}$, monitored by spectroscopic ellipsometry (SE). **Figure 2a** reports $n_e$ as a function of $n_D$ for all the SOI samples with $H_{SOI}$ = 30 ± 1 nm (black open symbols): the increase of $n_D$ directly correlates with the increase of $n_e$. The fraction of activated and ionized P dopant impurities ($\eta_a$) is computed as $\eta_a = n_e/n_D$ assuming $n_e$ is indicative of the ionized P at room temperature [21]. Accordingly, when $n_D > 1 \times 10^{18}$ cm$^{-3}$, $\eta_a$ values are nearly constant and well above 80 %. The theoretical model of incomplete ionization proposed by *Altermatt* et al. [33] predicts that in bulk Si, the fraction of ionized P impurity atoms changes with the concentration of the dopants. In particular, the occupation probability of dopant states is directly related to the relative position of the energy level of the dopant ($E_d$) with respect to the Fermi energy level ($E_F$). A minimum at $n_D \sim 2 \times 10^{18}$ atoms/cm$^3$ is expected because $E_F$ is close to $E_d$, and up to 25 % of donors are expected to be non-ionized resulting in significant incomplete ionization even at room temperature [33]. In this $n_D$ range, experimental data obtained when $H_{SOI} \sim 30$ nm are in excellent agreement with this model, confirming full activation of the dopants at room temperature [23]. Interestingly, $\eta_a \sim 55$ % is obtained for $n_D \sim 1 \times 10^{18}$ cm$^{-3}$. Considering that all the samples experienced the same annealing process, the fraction of active P atoms is expected to be the same. This reduction of $\eta_a$ suggests a decrease of $n_e$ well beyond the values expected according the incomplete ionization model proposed by *Altermatt* et al. [33]. This reduction will be investigated and discussed in detail in the following sections.

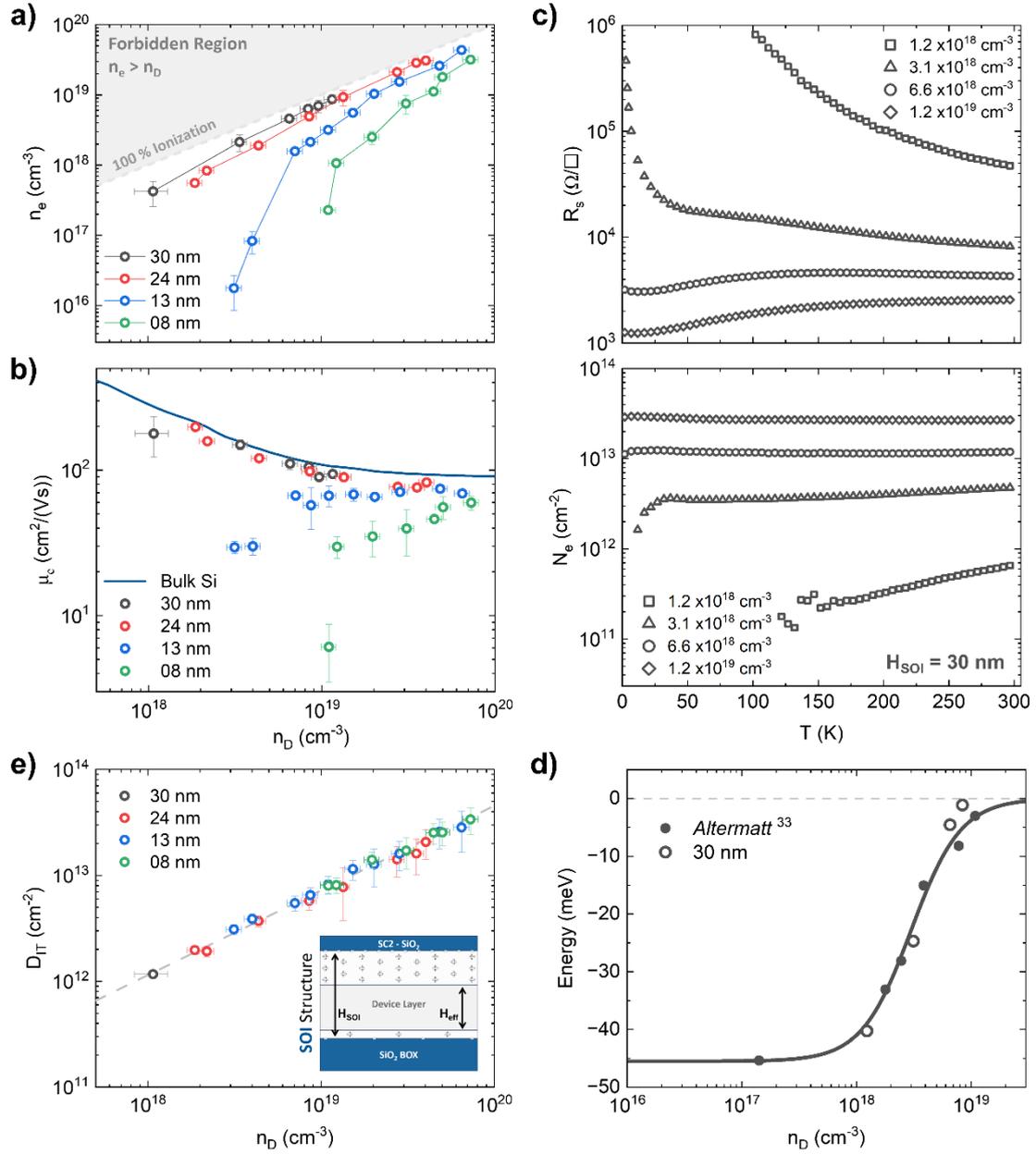

*Figure 2.* (a) Concentration of charge carriers ($n_e$) obtained by Hall measurements vs the total dopant concentration ($n_D$) measured by calibrated ToF-SIMS depth profiles for SOI with different $H_{SOI}$. (b) Mobility $\mu_e$ values as a function of $n_D$ for the same SOI samples. Electron mobility in bulk Si (line) is shown for comparison as reported by Sze [5]. (c) Example low temperature sheet resistance $R_s$ and total carrier dose $N_e$ obtained on SOI samples with $H_{SOI} \sim 30$ nm for different $n_D$ values below and above the metal-insulator transition. (d) Ionization energy obtained on SOI samples with $H_{SOI} = 30 \pm 1$ nm by eq. 2 versus $n_D$. The black line corresponds to the fitting function as eq. 3. Data obtained by Altermatt et al. are reported in the graph for comparison [33]. (e) $D_{IT}$ values of SC2-TOX/Si interfaces vs $n_D$ for SOI with different $H_{SOI}$ estimated using eq. 5. In the inset, schematic representation of the effect of trapped charge carries at the interfaces.

Under the assumption of a uniform P distribution throughout the entire Si device layer, the majority carrier mobility ($\mu_e$) in the device layer can be directly computed from the combination of independent sheet resistance and Hall measurements using the following equation:

$$\rho = (q\, n_e\, \mu_e)^{-1} \tag{1}$$

where $q$ is the electron charge. **Figure 2b** reports the computed $\mu_e$ values versus $n_e$. The $\mu_e$ values reported in the in the literature for bulk Si [5] are shown (solid blue line) for comparison. Experimental data confirms that the doped SOI samples with $H_{SOI} \sim 30$ nm perfectly match the electrical properties of a uniformly doped bulk Si substrate having the same $n_D$. This $\mu_e$ evolution suggests that the effective mass of the carriers remains consistent when reducing $H_{SOI} \sim 30$ nm. Interestingly, the sample doped with the lowest $n_D$, which resulted in lower $\eta_a$, exhibited a reduced mobility value. This mobility reduction is counterintuitive since a decrease of dopant concentration and ionization is expected to limit Coulomb scattering phenomena determining, in principle, an increase of carrier mobility. Further discussion about this mobility reduction will follow in the next sections.

The electrical properties of some of the doped SOI samples with $H_{SOI} \sim 30$ nm were accurately investigated through sheet resistance and Hall effect measurements in vdP configuration for temperatures ranging from 1.8 to 300 K. The four selected samples are characterized by $n_D$ values corresponding to $1.2 \times 10^{18}$, $3.1 \times 10^{18}$, $6.6 \times 10^{18}$ and $1.2 \times 10^{19}$ cm$^{-3}$. These specific P concentration values enable the possibility to investigate in these set of samples the transition from semiconductor to metal which, in bulk Si, is reported to occur at $n_D \sim 3 \times 10^{18}$ cm$^{-3}$ [33,34]. **Figure 2c** reports the evolution of $R_s$ (top) and $N_e$ (bottom) as a function of temperature for the selected samples. Interestingly, two totally divergent trends are observed: the two lowly doped samples are characterized by an increase of $R_s$ when decreasing the temperature, while the opposite evolution is observed in the two highly doped samples.

The strong increase of $R_s$ observed at low temperature in the sample with $n_D = 1.2 \times 10^{18}$ cm$^{-3}$ perfectly correlates with the progressive decrease of $N_e$, suggesting a typical semiconductor-like behavior with conductivity directly dependent on a thermally activated ionization process of the dopant impurities. In particular, below 100 K, the resistance increases with values above the detection limit of the instrument, because of the freeze-out of carriers that is expected to occur in doped Si with dopant concentration below the metal-semiconductor transition. The other resistive sample with $n_D = 3.1 \times 10^{18}$ cm$^{-3}$ offers a more complex picture, because $R_s$ and $N_e$ show only a quite limited evolution with temperature from 300 to 20 K. Below this threshold, $R_s$ progressively increases while $N_e$ drops down to smaller and smaller values, qualitatively resembling the evolution of the sample with $n_D = 1.2 \times 10^{18}$ cm$^{-3}$ and suggesting a shift of freeze-out temperature. This result can be easily explained on the basis of standard semiconductor physics: increasing $n_D$ and approaching the expected Mott transition value in bulk Si, the wavefunctions of the dopant atoms begin to overlap and form an energy band within the Si bandgap, lowering the energy gap between the dopant level and the conduction band [33]. The smaller energy gap requires less thermal energy for the dopants to be ionized and carrier freeze-out starts at lower temperatures than in lightly doped semiconductors.

Conversely in the two highly doped samples, a typical metal-like behavior is observed with $R_s$ progressively decreasing when reducing the temperature. At the same time, $N_e$ remains perfectly constant in the entire $T$ range, without any evidence of freeze-out, suggesting an overlap between the dopant energy band and the conduction band [33]. Accordingly, no increase in the resistivity is expected because of reduction of free carriers in the conduction band. Conversely, the reduction of $R_s$ at low temperatures perfectly correlates with the overall picture describing conduction in metals: decreasing $T$ determines a reduction of number of phonons resulting in increased mobility and lower resistivity in the electron gas [5,34].

Collected data clearly highlight that increasing $n_D$, the samples shift from semiconductor- to metal-like evolution of both $R_s$ and $N_e$. To further corroborate this picture, a twofold ionization energy model was used to fit the experimental data obtained in all the SOI samples with $H_{SOI} \sim 30$ nm that were measured, which exhibited a thermally activated process [34,35]:

$$\sigma = \frac{1}{\rho} = A\, exp\left(-\frac{E_1}{k_B T}\right) + B\, exp\left(-\frac{E_2}{k_B T}\right) \qquad (2)$$

Fitting the experimental data, two different ionization energies are obtained. **Figure S2** shows fitting curves of the samples with $n_D = 1.2 \times 10^{18}$ cm$^{-3}$ and $n_D = 3.1 \times 10^{18}$ cm$^{-3}$. The value of $E_1$ is associated to the P donor energy level $E_d$, while the value of $E_2$ is usually associated with the presence of the impurity band due to the high $n_D$ [33,36]. The values of $E_d$ computed for all the SOI samples measured at low temperature which exhibited thermally activated evolution of the conducting properties are reported versus $n_D$ in **Figure 2d**. The data obtained in this work are compared to those reported in the literature for similarly P doped bulk Si [33]. Experimental data were fitted following the equation that accounts for the reduction of the ionization energy expected for high doping concentration [33]:

$$E_d = \frac{E_{d,0}}{1+(\frac{n_D}{n_{ref}})^c} \qquad (3)$$

Where $E_{d,0}$ is the P donor energy level for diluted dopant concentrations and was assumed to be 45.5 meV [37] while $c = 2$. Accordingly, $n_{ref}$ was determined to be $(3.0 \pm 0.4) \times 10^{18}$ cm$^{-3}$. The value is perfectly compatible, within the experimental error, with the one determined by Altermatt in the case of a P doped bulk Si substrate [33]. Overall, when $H_{SOI} \sim 30$ nm, the experimental results were perfectly described by the model for bulk Si of *Altermatt et al.* supporting the idea that in this $n_D$ range, the 30 nm thick SOI samples have electrical characteristics that are almost perfectly equivalent to bulk doped silicon in agreement with our previous data [21,23].

Conversely, the situation is getting much more complex when reducing $H_{SOI}$ below 30 nm. **Figure 2a** also reports the results of electrical and compositional analysis in the case of SOI samples with $H_{SOI} = 24 \pm 1$, $13 \pm 1$ and $8.2 \pm 0.4$ nm. For these samples the $n_e$ values are always far away from the limit of 100 % ionization, with $\eta_a$ values quickly dropping to values well below 10 % when decreasing $n_D$. Qualitatively, the collected data clearly indicate that lower P dopant concentration and/or stronger 2D confinement result in significantly lower $n_e$ values. As discussed in a previous paragraph, a similar effect was observed in the 30 nm thick SOI samples with $n_D \sim 1 \times 10^{18}$ cm$^{-3}$, exhibiting a $n_e$ value that corresponded to an average $\eta_a \sim 55$ %. Interestingly, at high $n_D$ values, the effect is progressively reduced, and almost no signature of this reduction of carrier concentration is observed, even in the case of ultrathin SOI with $H_{SOI} \sim$ 13 and 8 nm. Assuming full activation of the dopants, these experimental results cannot be explained on the basis of the model of incomplete ionization proposed by Altermatt [33] indicating a clear departure from bulk Si characteristics. Further investigations are necessary to fully elucidate the origin of this reduction of carrier concentration in the channel.

**2.2 Interface defects and charge trapping.** In a previous publication, we observed that when $H_{SOI}$ is reduced below 30 nm, the model of incomplete ionization cannot be directly applied to ultrathin SOI substrates, because of the combined effect of the TOX and BOX interfaces [23]. In particular, as surface-to-volume ratio increases, conducting properties move from a bulk-like to an interface-driven behavior. At this level, the main interface effect to be considered is the presence of non-passivated interface state traps at the $Si/SiO_2$ interface between the Si device layer and the surrounding TOX and BOX films. These interface states are strictly related to the presence of different types of Si dangling bonds, like $P_{b0}$ and $P_{b1}$, which are typically located at a $Si/SiO_2$ interface with (100) orientation [38]. All type of dangling bonds are reported to be of amphoteric nature [39–41]. Their energy distribution comprises of two distinct peaks in the Si bandgap, located in the lower and upper half of the bandgap, which introduce donor-like and acceptor-like energy levels, respectively. In an *n*-type Si, the Fermi-level is located in the upper half of the bandgap, and the acceptor-like energy levels between the Fermi-level and the middle of the bandgap are negatively charged, while the ones above the Fermi-level are neutral. Negative mobile charges trapped in the dangling bonds at the $Si/SiO_2$ interface do not contribute to charge conduction and result in a depletion of mobile charges near the interface with the formation of an extended space charge region [42]. As a result, the effective thickness of the conductive channel ($H_{eff}$) is lower than the physical thickness $H_{SOI}$. As the density of interface states ($D_{IT}$) of the $Si/SiO_2$ interface increases, the extension of the depletion region progressively increases and $H_{eff}$ shrinks even further [23].

**Figure 2e** (inset) shows a schematic representation of the SOI structure elucidating the typical effect of the TOX and BOX interfaces in *n*-type-doped SOI. It is worth to note that, in our previous work, the proposed interpretation assumed perfectly equal and symmetric contribution coming from the two Si/SiO$_2$ interfaces [23]. Experimental data herein reported suggests that the most dominant contribution is to be attributed to the Si/TOX interface, as will be discussed in the next section. In general, the scheme of **figure 2e** assumes that the quality of the chemically grown SC2-TOX is worse than the pristine and as-fabricated BOX, leading to a larger depletion zone around the Si/TOX interface. It is important to note that the overall reduction of $H_{eff}$ is not affected by the effective distribution of the depletion layers in the device layer at the different interfaces. The intensity of the effect of interface states is directly related to both $n_D$ and $H_{SOI}$. Lower $n_D$ implies larger depletion regions to fully compensate for the $D_{IT}$. As shown in the case of the 30 nm thick SOI sample with $n_D$ ~ 1 × 10$^{18}$ cm$^{-3}$, a non-negligible contribution of the interfaces is clearly observed even in the case of thick SOI samples when considering sufficiently low dopant concentrations. Similarly, if $H_{SOI}$ is reduced, a proportionally bigger fraction of the volume of the device layer is depleted, resulting in a significant reduction of $n_e$ that could ultimately lead to a fully depleted Si device layer.

This reduction of carrier concentration in the channel strongly correlates with a degradation of the carrier mobility values, as reported in **Figure 2b** in the case of ultrathin SOI. Experimental data clearly suggest a relation with the dimensions of the depleted regions created by interface states. For a fixed $H_{SOI}$ value, the $\mu_e$ values are significantly reduced at low $n_D$ but they approach bulk mobility values at high $n_D$. Moreover, the mobility degradation is progressively enhanced when reducing the $H_{SOI}$ value. This effect cannot be explained considering scattering induced by interface roughness. Mobility degradation is clearly observed already in SOI samples with $H_{SOI}$ ~ 13 nm, while interface roughness is typically expected to play a major role only for $H_{SOI}$ < 5 nm [43]. Additionally, considering the samples with $H_{SOI}$ ~ 8 nm, $\mu_e$ is observed to increase as $n_D$ increases. This is totally counterintuitive since in bulk Si, higher $n_D$ values are associated with larger numbers of ionized dopants and consequently to a higher Coulomb scattering contribution determining a mobility degradation [5].

Accordingly, mobility evolution in ultrathin SOI substrates suggests a physical mechanism that is significantly different from the typical bulk-like one. Depleted regions are characterized by the presence of a space charge determined by the impurity ions that act as scattering centers for the electrons traveling in those regions. This impurity scattering contribution is greatly enhanced in those regions because Thomas-Fermi screening of ions is almost negligible [44,45]. Accordingly, higher $n_D$ values correspond to smaller depleted regions and higher screening. These considerations about mobility degradation further support the idea of the formation of extended depletion regions near the $SiO_2$/Si interfaces because of trapping of electrons at the interface.

In previous work we developed a simple electrostatic model to calculate the width of the depleted region [23]. The values of the total $D_{IT}$ which would result in such a reduction of carrier concentration in the channel, can be estimated by assuming charge neutrality of the interface and full depletion of the space charge region, i.e. assuming an abrupt transition between the depletion layer and the non-depleted semiconductor material. According to the incomplete ionization model of Altermatt the density of electrons $N_e^{Altermatt}$ in the device layer is directly connected to the active dopant concentration by the following equation:

$$N_e^{Altermatt} = n_D \, H_{SOI} \, f^{Altermatt}(n_D) \qquad (4)$$

where $f^{Altermatt}(n_D)$ is the donor ionization fraction [33]. The difference between the calculated $N_e^{Altermatt}$ value and the measured $N_e$ value indicates the amount of electrons trapped at interface states. Accordingly, we calculate the $D_{IT}$ as:

$$D_{IT} = \Delta N = N_e^{Altermatt} - N_e \qquad (5)$$

The calculated $D_{IT}$ values are plotted in **Figure 2e** as a function of $n_D$ for all the different $H_{SOI}$ values. Interestingly, the computed $D_{IT}$ values for $H_{SOI} \sim 24$, 13 and 8 nm follow the same experimental trend, suggesting a $H_{SOI}$ independent mechanism. This $D_{IT}$ evolution is further confirmed by the value (black open symbol) obtained in the case of the 30 nm thick SOI sample with $n_D \sim 1 \times 10^{18}$ cm$^{-3}$, which exhibits a significant reduction of carrier concentration even for higher $H_{SOI}$. The reduction of carrier concentration in the channel observed in ultrathin SOI substrates appears to be most significantly driven by the quality of the Si/SiO$_2$ interfaces and interface states, which are the same in all the samples, rather than on the thickness on the Si device layer. In addition, since we are considering highly doped Si substrates, the Fermi level is expected to be close to the conduction band minimum. Consequently, the calculated $D_{IT}$ values account for almost all the acceptor-like interface states in the upper half of the Si bandgap. It is important to note that the present $D_{IT}$ values have been calculated taking into account a contribution due to incomplete ionization of the dopant impurities that has been calculated according to Altermatt's model that was developed for bulk Si. The assumption that the model can be directly applied to ultrathin films is anything but trivial. Nevertheless, $D_{IT}$ values in this range were reported in the literature for similarly non-passivated Si/SiO$_2$ interfaces [46,47]. Moreover, higher $D_{IT}$ values are expected as $n_D$ is increased due to enhanced segregation of P dopants at the interfaces, potentially causing additional traps and doping-induced defects [48].

**2.3 Interface Characterization and Engineering.** To further corroborate this model and clarify the role of interface state, electrical paramagnetic resonance (EPR) characterization of the samples at cryogenic temperatures was performed, providing additional information regarding the ionization of the P dopants in the Si device layer, as well as the quality of the Si/TOX interface. EPR measurements were performed on three reference SOI samples with $H_{SOI}$ = 27.2 ± 0.8 nm and P concentration ranging from 1 to 5 × 10$^{18}$ cm$^{-3}$, as measured by calibrated ToF-SIMS depth profiles. The EPR spectra of these SOI samples are presented in **Figure 3a**. An explanatory data fitting is presented in **Figure 3b**, for the SOI sample with the highest P concentration of n$_D$ = 5.0 × 10$^{18}$ cm$^{-3}$. A main single line is detected at g = 1.9991 ± 0.0005 with a linewidth of 0.15 ± 0.02 mT. This resonant line is consistent with the presence of clusters of P dopants, considering the high n$_D$ value [49]. Moreover, the P line intensity is progressively reducing with the donors concentration, in accordance with the reduction of free carriers presented in **Figure 2a**. At higher g-factors, a broader signal is detected composed of different contributions, as highlighted by the data fitting. Si dangling bonds

(DB) resonant line is detected at a characteristic g-factors of g = 2.0056 ± 0.0005, as expected due to defects generated by wafer cutting [50]. In addition, the two detected contributions at g = 2.0030 ± 0.0005 and g = 2.0070 ± 0.0005 are consistent with the $P_{b0}$ centers at the Si/SiO$_2$ interface for the considered field orientation (B // [110]) [51,52].

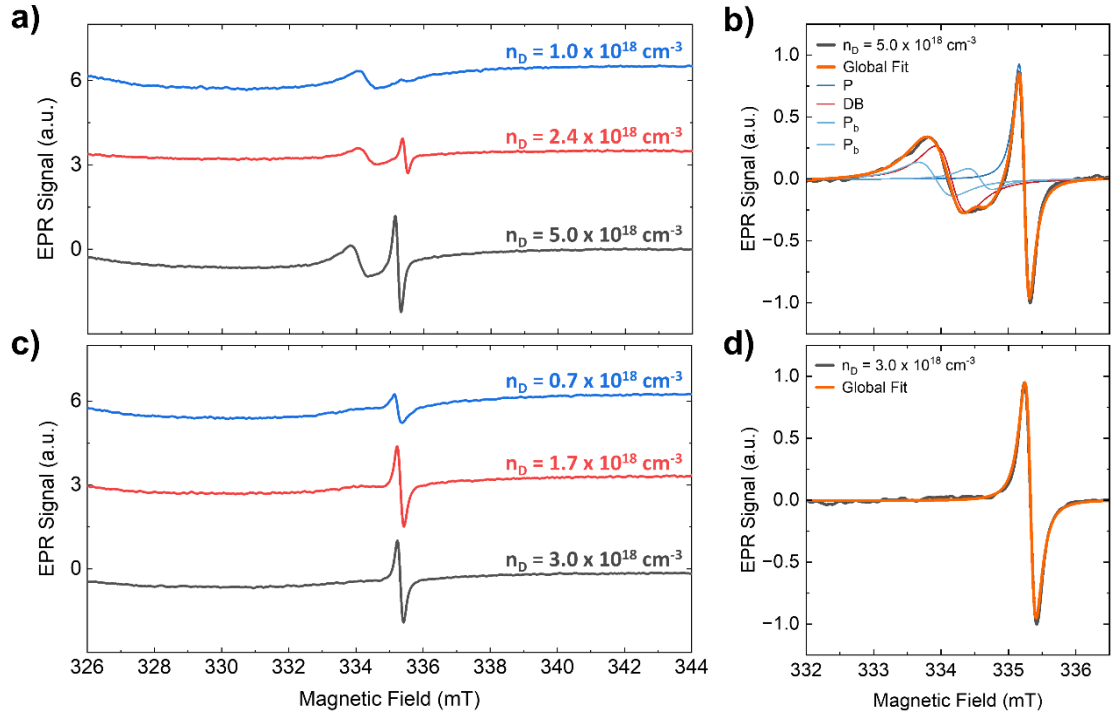

*Figure 3. (a) EPR spectrum of SOI samples with $H_{SOI}$ = 27.2 ± 0.8 nm doped with various P concentrations with SC2-TOX capping layer. (b) Fitting curve for the sample with $n_D$ = 5.0 × 10$^{18}$ cm$^{-3}$. The different contributions of the detected species are indicated: a single P line, and a broad line resulting from the superposition of the Si dangling bonds (DB) and interface defects ($P_{b0}$ centers) resonant lines. (c) EPR spectrum of the same SOI samples upon oxidation at T = 900 °C, showing the successful removal of all the interface defects signals. (d) Fitting curve for the sample with $n_D$ = 3.0 × 10$^{18}$ cm$^{-3}$ upon oxidation showing only the P single line. All EPR spectra were acquired at 4.2 K, with B // [011], and microwave power of 1 mW.*

The same samples underwent rapid thermal oxidation (RTO) at T = 900 °C for 40 s in O$_2$ atmosphere. A 5 nm thick SiO$_2$ TOX film was thermally grown on top of the Si device layer with a reduction of the Si device layer of about 2 nm. Upon RTO, the average $H_{SOI}$ is determined to be 25.4 ± 0.8 nm. The low annealing temperature is expected to limit P diffusion into the $SiO_2$ layers surrounding the Si device layer. Additionally, the SC2 layer was not removed prior to the thermal oxidation to prevent a significant out-diffusion of the P dopants from the Si surface during the initial stages of the oxidation process. Nevertheless, upon oxidation, all the samples exhibit a ~ 30 % reduction of P concentration, as highlighted by ToF-SIMS analysis. The EPR spectra of the SOI samples upon oxidation are presented in **Figure 3c** using the same color code and indicating the newly obtained $n_D$. It is possible to note that the P single line is not affected by the thermal treatment, while the broad signal, related to the silicon dangling bonds and interface defects, is removed upon oxidation, indicating a concentration of dandling bonds in the samples below the sensitivity limits of the system. This result is confirmed by the data fitting presented in **Figure 3d**, showing only the single P line at unvaried g = 1.9991 ± 0.0005 and linewidth of 0.17 ± 0.02 mT, within the experimental error. Even in the case of the SOI substrate with the lowest dopants concentration, the detection of a single line still suggests the presence of electrons delocalized in P clusters rather than isolated donors [49]. The P signal detected in the SOI sample with $n_D = 0.7 \times 10^{18}$ $cm^{-3}$ after oxidation is significantly higher than the one obtained in the pristine sample even if, during oxidation, a fraction of P atoms out-diffused from the Si device layer into the $SiO_2$ causing a significant $n_D$ reduction. A similar consideration applies for the sample with $n_D = 1.7 \times 10^{18}$ $cm^{-3}$. The EPR data are perfectly consistent with the model we proposed to account for reduction of carrier concentration in the channel: a lower $D_{IT}$ results in smaller depleted regions [23] and lower trapping of electrons at the interface determining higher $n_e$ values and, consequently, stronger P signal in the EPR spectra after RTO. Even more important, the absence of interface defects signal is a clear evidence of an improved interface quality. No significant improvements of the Si/BOX interface are expected to occur because of the RTO treatment. Accordingly, these data are indicative of a significant improvement of the Si/TOX interface.

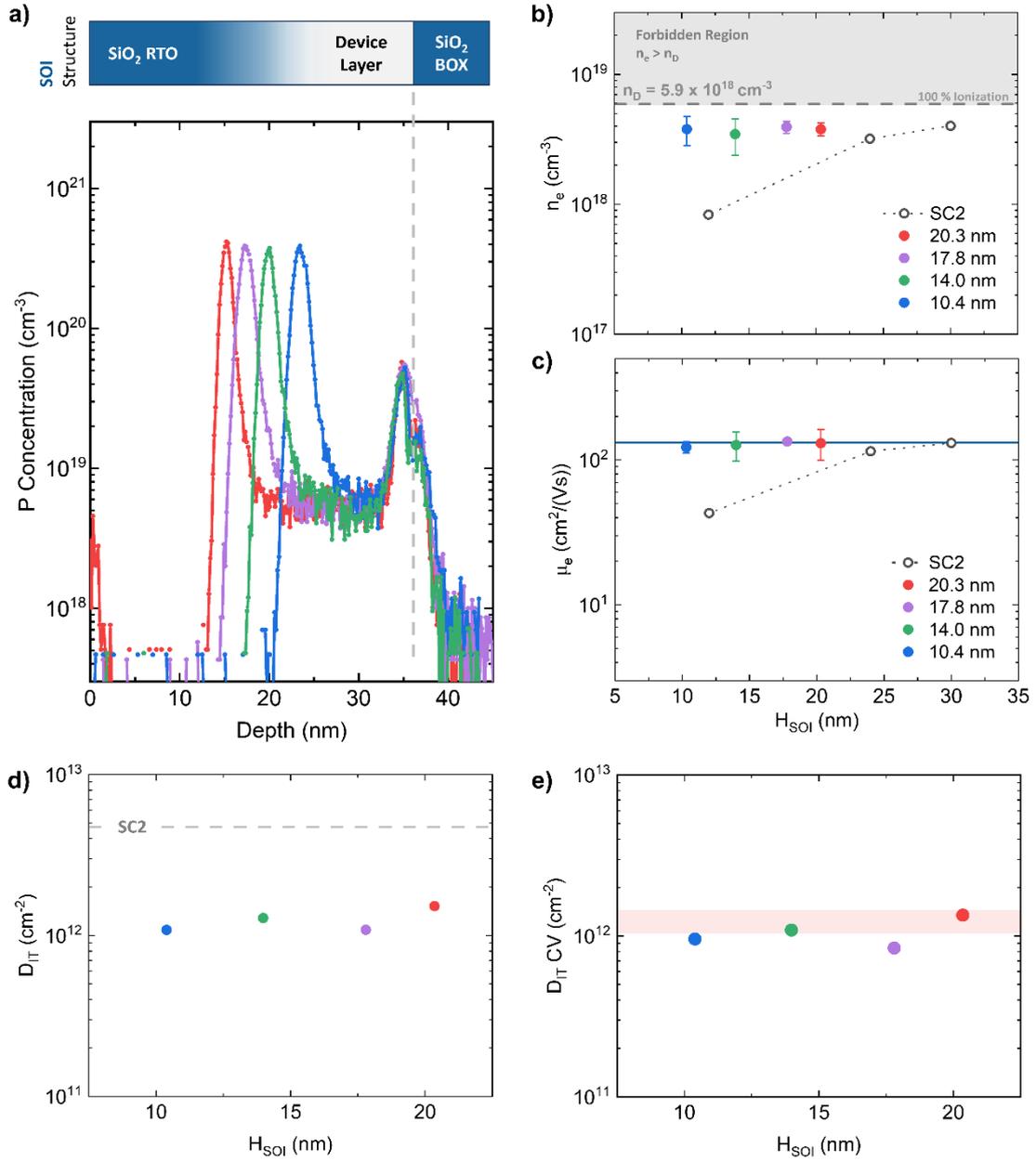

*Figure 4.* (a) ToF-SIMS depth profiles obtained on SOI samples upon RTO at $T = 900$ °C. The duration of the thermal treatment was selected to produce ultrathin films with $H_{SOI}$ ranging from 10.4 to 20.3 nm. (b) $n_e$ versus $H_{SOI}$ upon RTO vs SC2. Dashed line corresponds to the average $n_D$ value. (c) $\mu_e$ versus $H_{SOI}$ for the same set of samples. (d) $D_{IT}$ versus $H_{SOI}$ computed using eq. 5 for the same set of samples compared to the value expected for SOI samples with the same $n_D$ and SC2-TOX capping layer. (e) $D_{IT}$ values obtained by CV measurements versus $H_{SOI}$ for the same set of samples.

The EPR results suggest RTO is a valuable tool to engineer Si/TOX interface by reducing the trapping of electron at the interface and the depletion of the Si device layer that are assumed to determine the reduction of the charges in the Si channel. To support this claim, doped SOI samples with $H_{SOI}$ ~ 30 nm were prepared by means of a single grafting-hashing cycle using PS-P polymer (**Table S1**). Upon drive-in of the dopants, the samples were oxidized in $O_2$ atmosphere at $T = 900\,°C$ for different times progressively reducing the $H_{SOI}$ to ~ 20.3, 17.8, 14.0 and 10.4 nm. The calibrated ToF-SIMS P depth profiles of these samples after oxidation are reported in **Figure 4a**. The P depth profiles are shifted in depth to align the position of the Si/BOX interface. ToF-SIMS analysis demonstrates a uniform distribution of the P atoms into the Si device layer. No significant P signal is observed in the TOX and BOX films. Moreover, after oxidation, the dopant concentration is almost the same in all the samples irrespective of $H_{SOI}$. The average P concentration in the device layer was determined to be $n_D = (5.9 \pm 0.4) \times 10^{18}$ $cm^{-3}$ and is indicated as a grey dashed line in **Figure 4b**. Interestingly, the average $n_D$ expected after one doping cycle with this specific PS-P is $9 \times 10^{18}$ $cm^{-3}$, confirming a significant 35 % reduction of the P concentration after high temperature RTO. All the samples exhibit the same reduction of P concentration irrespective of the annealing time. Combining these ToF-SIMS data with those obtained in the case of samples used for the EPR analysis, we speculate that this reduction of P concentration is essentially related to a significant out-diffusion of the P atoms taking place during the initial stages of the RTO process, because of the low quality of the SC2-TOX film. Upon formation of a good quality thermal $SiO_2$ no further reduction of P concentration is observed.

**Figure 4b** shows the carrier concentration $n_e$ obtained from Hall measurements versus $H_{SOI}$ for all the four SOI samples upon oxidation. The results are compared to those (black open symbols) obtained on similarly doped SOI samples capped with a SC2-TOX layer. **Figure 4b** clearly shows a significant increase of $n_e$ in the samples with RTO-TOX capping layer. Moreover, no significant variation of the $n_e$ value is observed when reducing the thickness of the Si device layer. In particular, almost 10 times higher $n_e$ is observed in the 10.4 nm thick SOI sample with RTO-TOX capping layer compared to the 13 nm thick SOI samples with SC2-TOX capping layer. The $n_e$ values in the samples upon RTO processing indicated that, irrespective of the thickness of the device layer, an average $\eta_a$

~ 75 % was achieved in all the sample, within the limit of the theoretical model of incomplete ionization proposed by *Altermatt* et al. [33] Accordingly, almost no reduction of charge concentration is observed in the Si device layer of these samples even when reducing $H_{SOI}$ to values close to 10 nm. Considering the uniform P distribution in the Si device layers, the $\mu_e$ values in the different samples can be directly computed by means of **eq. 1** from the combination of separate sheet resistance and Hall measurements. **Figure 4c** reports the computed $\mu_e$ values versus $H_{SOI}$. The $\mu_e$ values are perfectly consistent with those reported in the literature for bulk Si (blue line) [5]. Mobility values (black open symbols) of SOI films with SC2-TOX capping layer having the same dopant concentration are reported for comparison. The collected data clearly highlight that the degradation of the mobility which was observed in **Figure 2b** when reducing the Si device layer thickness is completely recovered replacing the SC2-TOX capping layer with the RTO-TOX capping layer. These results further corroborate the assumption that the effective mass remains constant even for SOI samples with $H_{SOI}$ < 30 nm. All the collected data indicate interface defects as key element to explain the reduction of carrier concentration and the degradation of carrier mobility in ultra-thin Si device layers.

The values of the total $D_{IT}$ which result in the small reduction of the charge concentration into the Si device layer of the samples with RTO-TOX capping layer are estimated again by following the same protocol previously described in **eq. 5**. The computed $D_{IT}$ values are plotted in **Figure 4d** versus $H_{SOI}$. The average $D_{IT}$ value was determined to be $(1.2 \pm 0.2) \times 10^{12}$ cm$^{-2}$. Expected data are compared to the average $D_{IT}$ value computed in the case of sample with SC2-TOX capping layer and $n_D$ ~ 5.9 × 10$^{18}$ cm$^{-3}$. A reduction by a factor of 4 is clearly observed in the samples with RTO-TOX capping layer. Better quality of the thermally grown RTO-TOX film and consequently of the Si/TOX interface results in lower $D_{IT}$ values and lower trapping of charges at the interface states.

Taking advantage of the high quality of the RTO-TOX capping layer, CV measurements were performed onto simple MOS capacitors using square aluminum contacts deposited by thermal evaporation on top of the RTO-TOX layer. Back contact to the Si device layer was achieved by locally removing the RTO-TOX capping layer and depositing Al to form an ohmic contact with the Si device layer. The CV curves were acquired at room temperature, sweeping the applied voltage between inversion to accumulation at frequencies ranging from 1 kHz to 1 MHz in a dark environment in a shielded probe station. CV and conductance curves were analyzed to extract the average $D_{IT}$ value at the Si/TOX interface. The average $D_{IT}$ values extracted from analysis of the CV curves are reported in **Figure 4e** and compared to the average $D_{IT}$ value calculated combining $n_D$ and $n_e$ values obtained form ToF-SIMS analysis and electrical Hall measurements, respectively. Measured $D_{IT}$ values are perfectly compatible with those extracted from the electrical and compositional analysis. Interestingly, the $D_{IT}$ values measurements in SOI samples with the RTO-TOX capping layer are in perfect agreement, within the experimental error, with those measured in bulk Si with the same kind of RTO-TOX capping layer (**Figure S3**). It is worth to remember that the $D_{IT}$ extracted from the combination of the electrical and compositional analysis refers to the total amount of interface states which should contribute to trapping to account for the reduction of $n_e$. Conversely, the $D_{IT}$ extracted from CV measurements refers to only the interface states at the Si/TOX interface without considering the contribution of the Si/BOX interface. On the basis of the experimental results, we expect that the major contribution to interface states is from the high-quality Si/TOX interface which was significantly modified during the RTO treatment while the Si/BOX interface is essentially unaffected by the thermal treatment. For this reason, in the inset of **Figure 2e**, the contribution of the two interfaces is assumed to be significantly different with an important channel depletion in the proximity of the Si/TOX interface.

**2.5 Dielectric mismatch in ultrathin films.** Overall, data in **Figure 4** indicate that at room temperature the doped SOI samples with RTO-TOX capping layer behave like bulk Si even for samples with Si device layer thickness well below 30 nm. Actually, the situation is a bit more complex as highlighted by sheet resistance and Hall measurements at low temperature. **Figure 5** illustrates the low temperature evolution from 5 to 300 K of the sheet resistance and carrier concentration in the doped SOI films with RTO-TOX capping layer when $H_{SOI}$ is progressively reduced from 29.0 nm to 20.3, 14.0 and 10.4 nm. At $H_{SOI}$ ~30 nm the system behaves as a metallic material, in line with the expectation for bulk Si doped above the Mott transition. The average dopant concentration in the device layer is $(5.9 \pm 0.4) \times 10^{18}$ cm$^{-3}$. The carriers remain delocalized, and the conductivity shows only weak temperature dependence. Upon reducing $H_{SOI}$ to 20 nm, the metallicity begins to break down: a partial freeze-out of carriers is observed as temperature decreases below 25 K, suggesting that the effective donor ionization energy is already increasing in this regime. The trend becomes much more pronounced in the 14 and 10 nm thick films, where the conductivity is thermally activated and the carrier concentration essentially vanishes at low temperature, that is indicative of strong dopant freeze-out. The data indicate a shift of the P threshold concentration corresponding to Mott transition in agreement with data reported by Tanaka *et al*. [53]. The average ionization energy of the dopants for the samples with $H_{SOI}$ ~ 14.0 and 10.4 nm extracted from the Arrhenius behavior of the conductivity is $(51 \pm 3)$ meV, a value significantly larger than the one expected in similarly doped bulk Si, pointing to a fundamental modification of the donor energy in ultrathin SOI layers. Especially considering the reduction of $E_a$ which should be observed in bulk Si at high $n_D$ [33].

In general, surface phenomena cannot be discounted as thickness approaches and decreases below 10 nm. Many effects must be considered at these scales such as carrier surface scattering, dielectric screening [16], interface states [15], as well as decreased doping efficiency [10], increased dopant trapping [54], and an increase in Si bandgap [55]. It is important to note that such behavior cannot be explained by quantum confinement, since at $H_{SOI}$ ~ 10 nm the Si device layer is still much thicker than the Bohr radius of donors (~ 2-3 nm) and the electronic states are not expected to be strongly quantized in the conduction band. Instead, we assume that the dominant mechanism highlighted here is associated to the dielectric mismatch due to the surrounding SiO$_2$ [23]. The dielectric constant mismatch between Si ($\varepsilon_{Si} \approx 11.7$) and SiO$_2$ ($\varepsilon_{SiO2} \approx 3.9$) alters the Coulomb interaction of charged impurities: the lower dielectric constant of the oxide reduces the screening of the donor potential, thereby strengthening the Coulomb attraction and increasing the donor binding energy. This dielectric mismatch effect accounted for the strong increase of donor ionization energy that was reported in thin Si NWs embedded in SiO$_2$ [15]. Our data therefore provide direct experimental evidence of this dielectric confinement mechanism in ultrathin SOI. Despite the dopant concentration being above the threshold for Mott transition in bulk materials, carriers in the 10.4 and 14.0 nm thick layers become localized at low temperatures because of the enhanced ionization energy. This shift of the ionization energy explains the crossover from metallic behavior at 30 nm to semiconducting, activated behavior at 10.4 and 14.0 nm. The observed increase of ionization energy with decreasing thickness reflects the progressive strengthening of dielectric mismatch as the Si channel becomes thinner, in agreement with theoretical expectations [15,23].

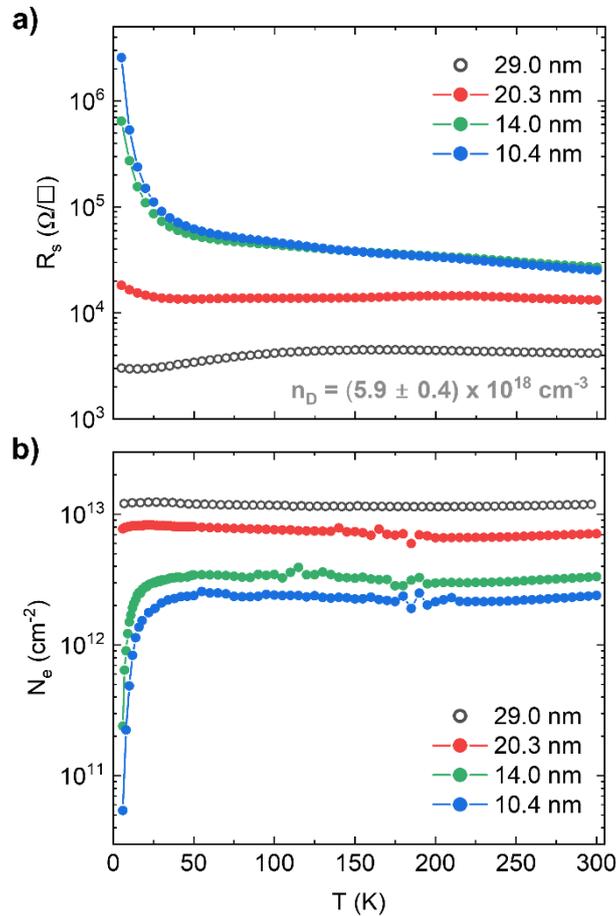

*Figure 5.* (a) low temperature sheet resistance $R_s$ and total carrier dose $N_e$ (b) obtained on SOI samples with different $H_{SOI}$ and $n_D = (5.9 \pm 0.4) \times 10^{18}$ cm$^{-3}$.

# 3. CONCLUSIONS

In this work, we examined the complex interplay between interface effects and electrical properties of ultrathin P-doped SOI films, focusing on the influence of non-passivated interface states and on the dielectric mismatch between Si and $SiO_2$. Varying $H_{SOI}$ from 8 to 30 nm and $n_D$ from $10^{18}$ to almost $10^{20}$ cm$^{-3}$, we explored a broad set of configurations, providing a comprehensive framework to assess the mechanisms of doping in these nanostructures. Our findings confirm that electrical properties are perfectly compatible with similarly doped bulk Si substrates for $H_{SOI} \sim 30$ nm, while a progressive degradation of electrical properties occurs as $H_{SOI}$ decreases below this threshold value.

At room temperature, progressive 2D confinement resulted in a progressively increasing reduction of charge carrier concentration and degradation of mobility. These phenomena are a direct result of two fundamental mechanisms that become dominant at the nanoscale: charge trapping at non-passivated interface states, which dominates at room temperature, and a significant increase in donor ionization energy caused by dielectric mismatch. Interface states act as charge traps, reducing the concentration of free carriers and creating a depletion layer near the Si/TOX interface that determines a mobility degradation due to increasing Coulomb scattering from the non-compensated ionized donors.

To reduce this effect, we performed an RTO annealing, which effectively improved the quality of the $SiO_2$/Si interface by reducing the $D_{IT}$. In this way we highlighted the influence of dielectric mismatch on the electrical characteristics of the P doped ultrathin SOI substrates. Low-temperature measurements of the samples with RTO-TOX capping layer revealed a significant increase in the P ionization energy for $H_{SOI} \leq 14.0$ nm, with an average extracted value of $(51 \pm 3)$ meV, as a direct consequence of dielectric confinement. In ultrathin films, charge carriers cannot effectively screen the Coulomb attraction between the donor ion and its valence electron because of the surrounding $SiO_2$ BOX and TOX layers, thereby increasing the binding energy of the dopants. Our findings provide direct experimental evidence of this effect on 2D confined Si films, which have been explored only in 1D systems like Si NWs.

In this respect, from a fundamental point of view, present results indicate that ultrathin SOI substrates could be exploited as an extremely advanced technological platform to better elucidate the complex interplay between different interface effects when changing the dopant concentration in spatially confined Si structures. Accordingly, when decreasing the Si device layer thickness well below 10 nm, these ultrathin SOI substrates could provide an extremely flexible and effective playground to investigate quantum confinement phenomena in a perfectly controlled system. From a technological point of view, our work shows that predictable and effective doping in ultrathin SOI for next-generation electronic devices requires precise dopant incorporation and interface engineering to account for the electrical properties of these nanostructures. Bulk doping models are insufficient at the nanoscale and need to be corrected to account for the role of interfaces. In conclusion, these findings provide critical insights into the fundamental physics governing dopant incorporation, activation, and ionization in a 2D confined environment and offer a new framework for designing and optimizing future SOI-based devices.

## 4. METHODS

**Substrate.** 1 × 1 cm² SOI dies were cleaved from lightly doped SOI wafers. Two distinct pristine SOI wafers from different suppliers were used to minimize the contribution of the substrate to the experimental results. The buried oxide (BOX) thickness of the SOI substrates was 200 and 160 nm, while the Si device layer thickness ($H_{SOI}$) was 75 and 50 nm, respectively. The device layer was thinned down following an oxidation procedure at $T = 1000$ °C that is fully described in our previous publications [21,23]. This process already demonstrated nanometric control on the thickness of ultrathin SOI films with no evidence of sample degradation associated with the high temperature oxidation [23].

**Doping protocol.** Polystyrene (PS) polymers end-terminated by a P-containing moiety (PS-P) were employed to create a P δ-layer source at the interface between the deglazed Si device layer and a 10 nm thick $SiO_2$ capping layer following a protocol fully described in previous publications [21,29].

PS-P is characterized by an average molar mass $M_n$ = 2.4 kg/mol and polydispersity index $Đ$ = 1.15. A self-limiting mechanism based on $M_n$ precisely determines the surface grafting density of the polymer in the brush layer and the P dose available in the δ-layer source [29]. The dopants are driven in via high temperature annealing in a rapid thermal processing (RTP) system. Higher grafting density already demonstrated higher dose of P dopants injected [56]. After removal of the $SiO_2$ capping layer, all the doped SOI samples were oxidized in SC2 solution ($H_2O:H_2O_2:HCl$, 5:1:1) at 75 °C for 20 min in order to guarantee a consistent $Si/SiO_2$ top oxide (TOX) interface. Finally, after mesa patterning of the samples in KOH solution (20 % wt.) at room temperature, aluminum metal contacts were deposited by thermal evaporation at the corners of each structure following two subsequent photolithography exposures.

**Characterization techniques.** Spectroscopic ellipsometry (SE) was used to accurately monitor $H_{SOI}$ after each step of the process, from oxidation to mesa patterning. Sample preparation was optimized to produce SOI films with $H_{SOI}$ ranging from 8 to 30 nm. The thickness of the device layer of each SOI sample is measured at nine evenly spaced points across the center of the sample at a fixed incidence angle of 75°. Repeated measurements of the same spot produced the same thickness value. An uncertainty of 0.1 nm was assigned to each measurement spot. The measurement scheme of an example sample is presented in **Figure S1**. Experimental data are fitted using a three-layer model comprising the BOX, the device layer, and the top $SiO_2$. Mean values and errors are computed as the average and standard deviation of the nine points collected across the sample. $H_{SOI}$ values reported in this paper in **Table S1** are the average ones obtained at the end of the process, after SC2 cleaning, just before mesa patterning and electrical characterization of the samples. **Table S1** contains the name and all the information regarding sample preparation details of all the SOI prepared from the two different substrates considered.

Sheet resistance and Hall measurements were carried out following the four-point probe (4PP) van der Pauw (vdP) method [57]. A constant current was injected through the Keithley 6221 current source, and the corresponding voltage was recorded with the Keithley 2182A nanovoltmeter. The

Newton-Raphson iterative method was used to solve the vdP formula numerically for $R_S$ [58]. During the Hall measurements, a magnetic field varying from + 0.8 to - 0.8 T was applied. All data were recorded and analyzed by a custom-written LabVIEW-based acquisition software. Accurate patterning of the substrates minimized the error produced in the measurements due to the displacement of contacts [59]. Six different vdP geometries, four squares and two crosses, 1.75 × 1.75 mm² each, were patterned at the center of each of the 1 × 1 cm² SOI and explored to reduce the contribution introduced by the finite size of the contacts [60]. Four device structures were contacted and measured on each SOI sample. Multiple measuring cycles were performed on each device. The results are calculated by averaging all the data recorded. To verify both the low-temperature conditions and the accuracy of the evaluation of the ionization energy of the dopants, a lightly P doped bulk Si ($\rho \sim$ 1-5 Ωcm) was characterized. $E_{d,1-5}$ was determined to be ~ 42 meV. The value is within experimental error compared to those reported for diluted P in the literature [37,61]. Low temperature sheet resistance and Hall measurements were carried out with the vdP option of DynaCool-12 system from Quantum Design. Sheet resistance was measured at zero magnetic field, while Hall measurements were performed at 3, 6, 7, and 8 T over a temperature range from 1.8 to 300 K.

ToF-SIMS measurements were performed in an IONTOF M6 system using $Cs^+$ ions for sputtering and $Bi^+$ ions for analysis. Depth scale calibration was performed by individually measuring the sputter rate in each of the SOI samples. The $Si/SiO_2$ interfaces were determined by the spikes in $^{30}Si$ signals, while $H_{SOI}$ was accurately determined by SE. To obtain quantitative information about the P concentration, the counts measured by the ToF-SIMS were converted into atom concentrations (atoms/cm³) following a calibration protocol fully described elsewhere [62]. The minimum concentration that can be discriminated due to the background of the detector in the measurement setup was directly measured in a low P doped bulk Si substrate ($\rho \sim$ 1-5 Ωcm). An uncertainty of about 10 % of the P concentration determined from SIMS analysis was attributed to each measurement. When multiple SOI samples with similar $H_{SOI}$ and $n_D$ are considered, experimental results are presented as the average of multiple measurements taken across different samples.

Continuous wave electron paramagnetic resonance (cw-EPR) measurements were performed in a Varian E15 EPR spectrometer equipped with a Bruker super high-Q cylindrical cavity (EF4122sHQ) resonating in the X-band (≈ 9.4 GHz). A microwave power of 1 mW was employed. A static magnetic field was applied parallel to the sample [011] direction and modulated, using an external Stanford SR830 lock-in amplifier, with a frequency of 100 kHz and amplitude of 0.1 mT. The g-factors were calibrated with a 2,2-diphenyl-1-picrylhydrazyl (DPPH, $g$ = 2.0036 ± 0.0003) standard. Measurements were performed at 4.2 K, using an Oxford instrument ESR900 helium flow cryostat. EPR spectra were recorded with a custom-written LabVIEW based acquisition software and data fitting performed with EasySpin MATLAB toolbox [63].

## ASSOCIATED CONTENT

### Supporting Information

**Table S1** contains all the relevant sample preparation data, including the name and characteristics of each SOI sample prepared. **Figure S1** presents the spectroscopic ellipsometry measurements scheme of an example SOI sample. **Figure S2** presents the fitting curves of the samples with $H_{SOI}$ ~ 30 nm and $n_D$ = 1.2 × 10$^{18}$ cm$^{-3}$ and $n_D$ = 3.1 × 10$^{18}$ cm$^{-3}$. **Figure S3** presents the $D_{IT}$ values obtained by CV analysis on bulk Si substrates oxidized following the same protocol used for the SOI samples, which validates the $D_{IT}$ values reported in the article.

### Data Availability

The data that support the findings of this study are available from the corresponding author upon reasonable request.

### AUTHOR INFORMATION

**Corresponding Author:** andreapulici@cnr.it, gabriele.seguini@cnr.it, marco.fanciulli@unito.it, michele.perego@cnr.it

# AUTHOR CONTRIBUTIONS

**Andrea Pulici**: Writing – original draft, Investigation, Formal analysis, Data curation.

**Gabriele Seguini**: Writing – review & editing, Investigation, Conceptualization.

**Fabiana Taglietti**: Writing – review & editing, Investigation, Formal analysis, Data curation.

**Roman Gumeniuk**: Writing – review & editing, Investigation, Data curation.

**Riccardo Chiarcos**: Writing – review & editing.

**Michele Laus**: Writing – review & editing, Supervision.

**Johannes Heitmann**: Writing – review & editing, Supervision.

**Marco Fanciulli**: Writing – review & editing, Supervision.

**Michele Perego**: Writing – original draft, Supervision, Formal analysis, Data curation, Conceptualization.

# Notes

The authors have no conflicts to disclose.

# Acknowledgements

This work was financially supported by the Italian project DONORS (grant number 2022WBPHKF). The project received funding from the Italian programme for Research Projects of National Interest (PRIN) in the framework of the National Recovery and Resilience Plan (PNRR). The DynaCool-12 measurement system was acquired by the DFG grant 422219907.

# REFERENCES

1 W. Shockley, *Bell Syst. Tech. J.*, 1949, **28**, 435–489.
2 I. Capan, B. Pivac and R. Slunjski, *Phys. Status Solidi C*, 2011, **8**, 816–818.
3 A. Rogalski, *Infrared Phys. Technol.*, 2002, **43**, 187–210.
4 C. Zhang, S. Chang and Y. Dan, *Adv. Phys. X*, 2021, **6**, 1871407.
5 S. M. Sze and K. K. Ng, *Physics of Semiconductor Devices*, John Wiley & Sons, Inc., Hoboken, NJ, USA, 2006.
6 H.-J. Gossmann and E. F. Schubert, *Crit. Rev. Solid State Mater. Sci.*, 1993, **18**, 1–67.
7 M. Y. Simmons, F. J. Ruess, K. E. J. Goh, T. Hallam, S. R. Schofield, L. Oberbeck, N. J. Curson, A. R. Hamilton, M. J. Butcher, R. G. Clark and T. C. G. Reusch, *Mol. Simul.*, 2005, **31**, 505–515.
8 J. Foggiato and W. S. Yoo, *J. Vac. Sci. Technol. B Microelectron. Nanometer Struct. Process. Meas. Phenom.*, 2006, **24**, 515–520.
9 J. C. Ho, R. Yerushalmi, Z. A. Jacobson, Z. Fan, R. L. Alley and A. Javey, *Nat. Mater.*, 2008, **7**, 62–67.
10 T.-L. Chan, M. L. Tiago, E. Kaxiras and J. R. Chelikowsky, *Nano Lett.*, 2008, **8**, 596–600.
11 M. Perego, C. Bonafos and M. Fanciulli, *Nanotechnology*, 2009, **21**, 025602.
12 B. L. Oliva-Chatelain, T. M. Ticich and A. R. Barron, *Nanoscale*, 2016, **8**, 1733–1745.
13 I. Marri, E. Degoli and S. Ossicini, *Prog. Surf. Sci.*, 2017, **92**, 375–408.
14 G. Seguini, C. Castro, S. Schamm-Chardon, G. BenAssayag, P. Pellegrino and M. Perego, *Appl. Phys. Lett.*, 2013, **103**, 023103.
15 M. T. Björk, H. Schmid, J. Knoch, H. Riel and W. Riess, *Nat. Nanotechnol.*, 2009, **4**, 103–107.
16 M. Diarra, Y.-M. Niquet, C. Delerue and G. Allan, *Phys. Rev. B*, 2007, **75**, 045301.
17 D. König, D. Hiller, S. Gutsch, M. Zacharias and S. Smith, *Sci. Rep.*, 2017, **7**, 46703.
18 R. J. Theeuwes, W. M. M. Kessels and B. Macco, *J. Vac. Sci. Technol. A*, 2024, **42**, 060801.
19 B. Weber, S. Mahapatra, H. Ryu, S. Lee, A. Fuhrer, T. C. G. Reusch, D. L. Thompson, W. C. T. Lee, G. Klimeck, L. C. L. Hollenberg and M. Y. Simmons, *Science*, 2012, **335**, 64–67.
20 G. K. Celler and S. Cristoloveanu, *J. Appl. Phys.*, 2003, **93**, 4955–4978.
21 A. Pulici, S. Kuschlan, G. Seguini, F. Taglietti, M. Fanciulli, R. Chiarcos, M. Laus and M. Perego, *Mater. Sci. Semicond. Process.*, 2023, **163**, 107548.
22 F. Gity, F. Meaney, A. Curran, P. K. Hurley, S. Fahy, R. Duffy and L. Ansari, *J. Appl. Phys.*, 2021, **129**, 015701.
23 A. Pulici, S. Kuschlan, G. Seguini, M. D. Michielis, R. Chiarcos, M. Laus, M. Fanciulli and M. Perego, *J. Mater. Chem. C*, 2024, **12**, 18772–18778.
24 C. Ahn, N. Bennett, S. T. Dunham and N. E. B. Cowern, *Phys. Rev. B*, 2009, **79**, 073201.
25 N. Chery, M. Zhang, R. Monflier, N. Mallet, G. Seine, V. Paillard, J. M. Poumirol, G. Larrieu, A. S. Royet, S. Kerdilès, P. Acosta-Alba, M. Perego, C. Bonafos and F. Cristiano, *J. Appl. Phys.*, 2022, **131**, 065301.
26 S. Kerdilès, M. Opprecht, D. Bosch, M. Ribotta, B. Sklénard, L. Brunet and P. P. Michalowski, *Mater. Sci. Semicond. Process.*, 2025, **186**, 109043.
27 B. Ghyselen, J.-M. Hartmann, T. Ernst, C. Aulnette, B. Osternaud, Y. Bogumilowicz, A. Abbadie, P. Besson, O. Rayssac, A. Tiberj, N. Daval, I. Cayrefourq, F. Fournel, H. Moriceau, C. Di Nardo, F. Andrieu, V. Paillard, M. Cabié, L. Vincent, E. Snoeck, F. Cristiano, A. Rocher, A. Ponchet, A. Claverie, P. Boucaud, M.-N. Semeria, D. Bensahel, N. Kernevez and C. Mazure, *Solid-State Electron.*, 2004, **48**, 1285–1296.
28 A. Ohata, N. Rodriguez, C. Navarro, L. Donetti, F. Gamiz, F. C. Fenouillet-Beranger and S. Cristoloveanu, *J. Appl. Phys.*, 2013, **113**, 144514.
29 M. Perego, G. Seguini, E. Arduca, A. Nomellini, K. Sparnacci, D. Antonioli, V. Gianotti and M. Laus, *ACS Nano*, 2018, **12**, 178–186.
30 M. Perego, F. Caruso, G. Seguini, E. Arduca, R. Mantovan, K. Sparnacci and M. Laus, *J. Mater. Chem. C*, 2020, **8**, 10229–10237.
31 M. Perego, G. Seguini, E. Mascheroni, E. Arduca, V. Gianotti and M. Laus, *J. Mater. Chem. C*, 2021, **9**, 4020–4028.
32 A. Fick, *Ann. Phys.*, 1855, **170**, 59–86.
33 P. P. Altermatt, A. Schenk and G. Heiser, *J. Appl. Phys.*, 2006, **100**, 113714.
34 N. F. Mott and W. D. Twose, *Adv. Phys.*, 1961, **10**, 107–163.

35 H. Fritzsche, *Phys. Rev.*, 1955, **99**, 406–419.
36 P. R. Cullis and J. R. Marko, *Phys. Rev. B*, 1975, **11**, 4184–4200.
37 B. Pajot, J. Kauppinen and R. Anttila, *Solid State Commun.*, 1979, **31**, 759–763.
38 C. R. Helms and E. H. Poindexter, *Rep. Prog. Phys.*, 1994, **57**, 791.
39 M. Xiao, I. Martin and H. W. Jiang, *Phys. Rev. Lett.*, 2003, **91**, 078301.
40 M. Xiao, I. Martin, E. Yablonovitch and H. W. Jiang, *Nature*, 2004, **430**, 435–439.
41 J. T. Ryan, P. M. Lenahan, A. T. Krishnan and S. Krishnan, *J. Appl. Phys.*, 2010, **108**, 064511.
42 V. Schmidt, S. Senz and U. Gösele, *Appl. Phys. A*, 2007, **86**, 187–191.
43 G. Iannaccone, F. Bonaccorso, L. Colombo and G. Fiori, *Nat. Nanotechnol.*, 2018, **13**, 183–191.
44 N. W. Ashcroft and N. D. Mermin, *Solid State Physics*, Cengage Learning, 2011.
45 S. Das Sarma and E. H. Hwang, *Sci. Rep.*, 2015, **5**, 16655.
46 H. Angermann, Th. Dittrich and H. Flietner, *Appl. Phys. A*, 1994, **59**, 193–197.
47 W. Lu, C. Leendertz, L. Korte, J. A. Töfflinger and H. Angermann, *Energy Procedia*, 2014, **55**, 805–812.
48 J. Snel, *Solid-State Electron.*, 1981, **24**, 135–139.
49 G. Feher, *Phys. Rev.*, 1959, **114**, 1219–1244.
50 J. L. Cantin and H. J. von Bardeleben, *J. Non-Cryst. Solids*, 2002, **303**, 175–178.
51 P. M. Lenahan and J. F. Conley, *J. Vac. Sci. Technol. B Microelectron. Nanometer Struct. Process. Meas. Phenom.*, 1998, **16**, 2134–2153.
52 A. Stesmans and V. V. Afanas'ev, *J. Appl. Phys.*, 1998, **83**, 2449–2457.
53 T. Tanaka, Y. Kurosawa, N. Kadotani, T. Takahashi, S. Oda and K. Uchida, *Nano Lett.*, 2016, **16**, 1143–1149.
54 M. V. Fernández-Serra, Ch. Adessi and X. Blase, *Phys. Rev. Lett.*, 2006, **96**, 166805.
55 L. Lin, Z. Li, J. Feng and Z. Zhang, *Phys. Chem. Chem. Phys.*, 2013, **15**, 6063–6067.
56 M. Perego, S. Kuschlan, G. Seguini, R. Chiarcos, V. Gianotti, D. Antonioli, K. Sparnacci and M. Laus, *ACS Appl. Polym. Mater.*, 2021, **3**, 6383–6393.
57 L. J. van der PAUW, in *Semiconductor Devices: Pioneering Papers*, WORLD SCIENTIFIC, 1991, pp. 174–182.
58 In *Statistical Methods for Survival Data Analysis*, John Wiley & Sons, Ltd, 2003, pp. 428–432.
59 D. W. Koon, *Rev. Sci. Instrum.*, 1989, **60**, 271–274.
60 M. Reveil, V. C. Sorg, E. R. Cheng, T. Ezzyat, P. Clancy and M. O. Thompson, *Rev. Sci. Instrum.*, 2017, **88**, 094704.
61 A. Debernardi, A. Baldereschi and M. Fanciulli, *Phys. Rev. B*, 2006, **74**, 035202.
62 M. Mastromatteo, E. Arduca, E. Napolitani, G. Nicotra, D. De Salvador, L. Bacci, J. Frascaroli, G. Seguini, M. Scuderi, G. Impellizzeri, C. Spinella, M. Perego and A. Carnera, *Surf. Interface Anal.*, 2014, **46**, 393–396.
63 S. Stoll and A. Schweiger, *J. Magn. Reson.*, 2006, **178**, 42–55.

# Supporting Information

# Interface effects and dielectric mismatch in ultrathin silicon on insulator films

Andrea Pulici [a,b,*], Gabriele Seguini [a,*], Fabiana Taglietti [b], Roman Gumeniuk [c], Riccardo Chiarcos [d], Michele Laus [d], Johannes Heitmann [e], Marco Fanciulli [f], Michele Perego [a,*]

[a] CNR-IMM, Unit of Agrate Brianza, Via C. Olivetti 2, 20864 Agrate Brianza, Italy
[b] Università degli Studi di Milano-Bicocca, Via Roberto Cozzi 55, 20125 Milano, Italy
[c] Institut für Experimentelle Physik, TU Bergakademie Freiberg, Germany
[d] Università del Piemonte Orientale ''A. Avogadro'', Viale T. Michel 11, 15121 Alessandria, Italy
[e] Institut für Angewandte Physik, TU Bergakademie Freiberg, Germany
[f] University of Torino, Department of Chemistry, Via P. Giuria 9, 10125 Torino, Italy

[*]Corresponding Author:
andreapulici@cnr.it, gabriele.seguini@cnr.it, marco.fanciulli@unito.it, michele.perego@cnr.it

## Contents

1. ***Table S1: Sample preparation data***

2. ***Figure S1: Spectroscopic Ellipsometry (SE) measurements scheme***

3. ***Figure S2: Conductivity fit of the samples with $H_{SOI} \sim 30$ nm***

4. ***Figure S3: Capacitance-Voltage (CV) measurements on Bulk Si***

# 1. *Table S1: Sample preparation data*

**Substrate A** ($H_{SOI}$ = 75 nm, $H_{BOX}$ = 200 nm)

| SAMPLE | $H_{SOI}$ nm | Grafting # Cycles | Double Annealing | 1st Anneal T (°C) | t (s) | Capping | $n_D$ cm$^{-3}$ | $N_e$ cm$^{-2}$ |
|---|---|---|---|---|---|---|---|---|
| **30-3** | 30.3 | 1 | Y | 1000 | 10 | SC2 | 3.6E+18 | 7.72E+12 |
| **30-6** | 30.4 | 1 | N | --- | --- | SC2 | 8.4E+18 | 1.92E+13 |
| **30-6** | 31.2 | 1 | N | --- | --- | SC2 | 9.7E+18 | 1.87E+13 |
| **30-6** | 30.5 | 1 | N | --- | --- | SC2 | 9.7E+18 | 1.92E+13 |
| **24-1** | 22.8 | 1 | Y | 1000 | 1 | SC2 | 1.9E+18 | 1.27E+12 |
| **24-2** | 23.0 | 1 | Y | 1000 | 3 | SC2 | 2.2E+18 | 1.93E+12 |
| **24-3** | 23.5 | 1 | Y | 1000 | 10 | SC2 | 4.4E+18 | 4.48E+12 |
| **24-4** | 23.2 | 1 | Y | 1000 | 30 | SC2 | 8.4E+18 | 1.04E+13 |
| **24-5** | 23.2 | 1 | Y | 1000 | 100 | SC2 | 8.6E+18 | 1.26E+13 |
| **24-6** | 22.8 | 1 | N | --- | --- | SC2 | 1.5E+19 | 2.31E+13 |
| **24-6** | 26.2 | 1 | N | --- | --- | SC2 | 1.3E+19 | 2.15E+13 |
| **24-7** | 24.7 | 3 | N | --- | --- | SC2 | 1.3E+19 | 3.02E+13 |
| **24-8** | 24.9 | 5 | N | --- | --- | SC2 | 2.7 E+19 | 5.26E+13 |
| **24-8** | 23.0 | 5 | N | --- | --- | SC2 | 4.0E+19 | 7.10E+13 |
| **24-9** | 24.9 | 10 | N | --- | --- | SC2 | 3.6 E+19 | 7.11E+13 |
| **13-1** | 12.7 | 1 | Y | 1000 | 1 | SC2 | 3.5E+18 | 2.25E+10 |
| **13-2** | 12.5 | 1 | Y | 1000 | 3 | SC2 | 4.0E+18 | 1.04E+11 |
| **13-3** | 12.1 | 1 | Y | 1000 | 10 | SC2 | 7.7E+18 | 1.91E+12 |
| **13-4** | 12.0 | 1 | Y | 1000 | 30 | SC2 | 8.7E+18 | 2.57E+12 |
| **13-5** | 12.0 | 1 | Y | 1000 | 100 | SC2 | 1.2E+19 | 3.81E+12 |
| **13-6** | 11.5 | 1 | N | --- | --- | SC2 | 1.4E+19 | 4.87E+12 |
| **13-6** | 14.8 | 1 | N | --- | --- | SC2 | 1.6E+19 | 1.02E+13 |
| **13-7** | 11.9 | 3 | N | --- | --- | SC2 | 2.9E+19 | 1.53E+13 |
| **13-8** | 14.4 | 5 | N | --- | --- | SC2 | 2.8E+19 | 2.58E+13 |
| **13-8** | 11.8 | 5 | N | --- | --- | SC2 | 4.8E+19 | 3.07E+13 |
| **13-9** | 13.5 | 10 | N | --- | --- | SC2 | 6.5E+19 | 5.90E+13 |
| **08-1** | 8.5 | 1 | Y | 1000 | 1 | SC2 | 2.3E+18 | --- |
| **08-2** | 7.8 | 1 | Y | 1000 | 3 | SC2 | 4.5E+18 | --- |
| **08-3** | 8.1 | 1 | Y | 1000 | 10 | SC2 | 7.9E+18 | --- |
| **08-4** | 8.0 | 1 | Y | 1000 | 30 | SC2 | 1.1E+19 | 1.84E+11 |
| **08-5** | 8.5 | 1 | Y | 1000 | 100 | SC2 | 1.2E+19 | 9.09E+11 |
| **08-6** | 8.2 | 1 | N | --- | --- | SC2 | 2.0E+19 | 1.27E+12 |
| **08-6** | 9.0 | 1 | N | --- | --- | SC2 | 1.9E+19 | 2.58E+12 |
| **08-7** | 8.3 | 3 | N | --- | --- | SC2 | 2.9E+19 | 6.53E+12 |
| **08-8** | 8.0 | 5 | N | --- | --- | SC2 | 5.0E+19 | 1.44E+13 |
| **08-8** | 7.6 | 5 | N | --- | --- | SC2 | 4.5E+19 | 8.54E+12 |
| **08-9** | 8.3 | 10 | N | --- | --- | SC2 | 7.3E+19 | 2.62E+13 |
| **20-R** | 20.3 | 1 | N | --- | --- | RTO | 5.6E+18 | 8.40E+12 |
| **18-R** | 17.8 | 1 | N | --- | --- | RTO | 5.8E+18 | 7.38E+12 |
| **14-R** | 14.0 | 1 | N | --- | --- | RTO | 5.6E+18 | 5.82E+12 |
| **10-R** | 10.3 | 1 | N | --- | --- | RTO | 6.1E+18 | 4.95E+12 |

**Substrate B** ($H_{SOI}$ = 50 nm, $H_{BOX}$ = 160 nm)

| SAMPLE | $H_{SOI}$ | Grafting | Double Annealing | 1st Anneal | | Capping | $n_D$ | $N_e$ |
|---|---|---|---|---|---|---|---|---|
| | nm | # Cycles | | T (°C) | t (s) | | cm$^{-3}$ | cm$^{-2}$ |
| **30-1** | 31.9 | 1 | Y | 1000 | 1 | SC2 | 9.1E+17 | 1.73E+12 |
| **30-1** | 27.1 | 1 | Y | 1000 | 1 | SC2 | 1.2E+18 | 8.31E+11 |
| **30-3** | 29.1 | 1 | Y | 1000 | 10 | SC2 | 3.1E+18 | 5.04E+12 |
| **30-5** | 29.0 | 1 | Y | 1000 | 100 | SC2 | 6.6E+18 | 1.33E+13 |
| **30-6** | 31.5 | 1 | N | --- | --- | SC2 | 9.5E+18 | 2.70E+13 |
| **30-6** | 29.0 | 1 | N | --- | --- | SC2 | 1.2E+19 | 2.52E+13 |
| **24-6** | 24.5 | 1 | N | --- | --- | SC2 | 1.3E+19 | 1.67E+13 |
| **13-6** | 14.2 | 1 | N | --- | --- | SC2 | 2.0E+19 | 1.47E+13 |
| **08-6** | 8.9 | 1 | N | --- | --- | SC2 | 3.3E+19 | 6.32E+12 |

| SAMPLE EPR | $H_{SOI}$ | Grafting | Double Annealing | 1st Anneal | | Capping | $n_D$ | $N_e$ |
|---|---|---|---|---|---|---|---|---|
| | nm | # Cycles | | T (°C) | t (s) | | cm$^{-3}$ | cm$^{-2}$ |
| **E27-1** | 27.9 | 1 | Y | 900 | 10 | SC2 | 1.0E+18 | 1.52E+11 |
| **E27-2** | 27.8 | 1 | Y | 900 | 100 | SC2 | 2.4E+18 | 2.05E+12 |
| **E27-3** | 26.6 | 1 | Y | 900 | 300 | SC2 | 5.0E+18 | 6.73E+12 |
| **E27-1-R** | 26.0 | 1 | Y | 900 | 10 | RTO | 7.0E+17 | --- |
| **E27-2-R** | 25.7 | 1 | Y | 900 | 100 | RTO | 1.7E+18 | --- |
| **E27-3-R** | 24.4 | 1 | Y | 900 | 300 | RTO | 3.0E+18 | --- |

*Table S1. Summary of sample preparation and device characteristics of the SOI samples. The table details the name and the thickness of the samples, number of grafting cycles, the double annealing process (including first anneal parameters), the capping layer, and key measured parameters.*

## 2. *Figure S1: Spectroscopic Ellipsometry (SE) measurements scheme*

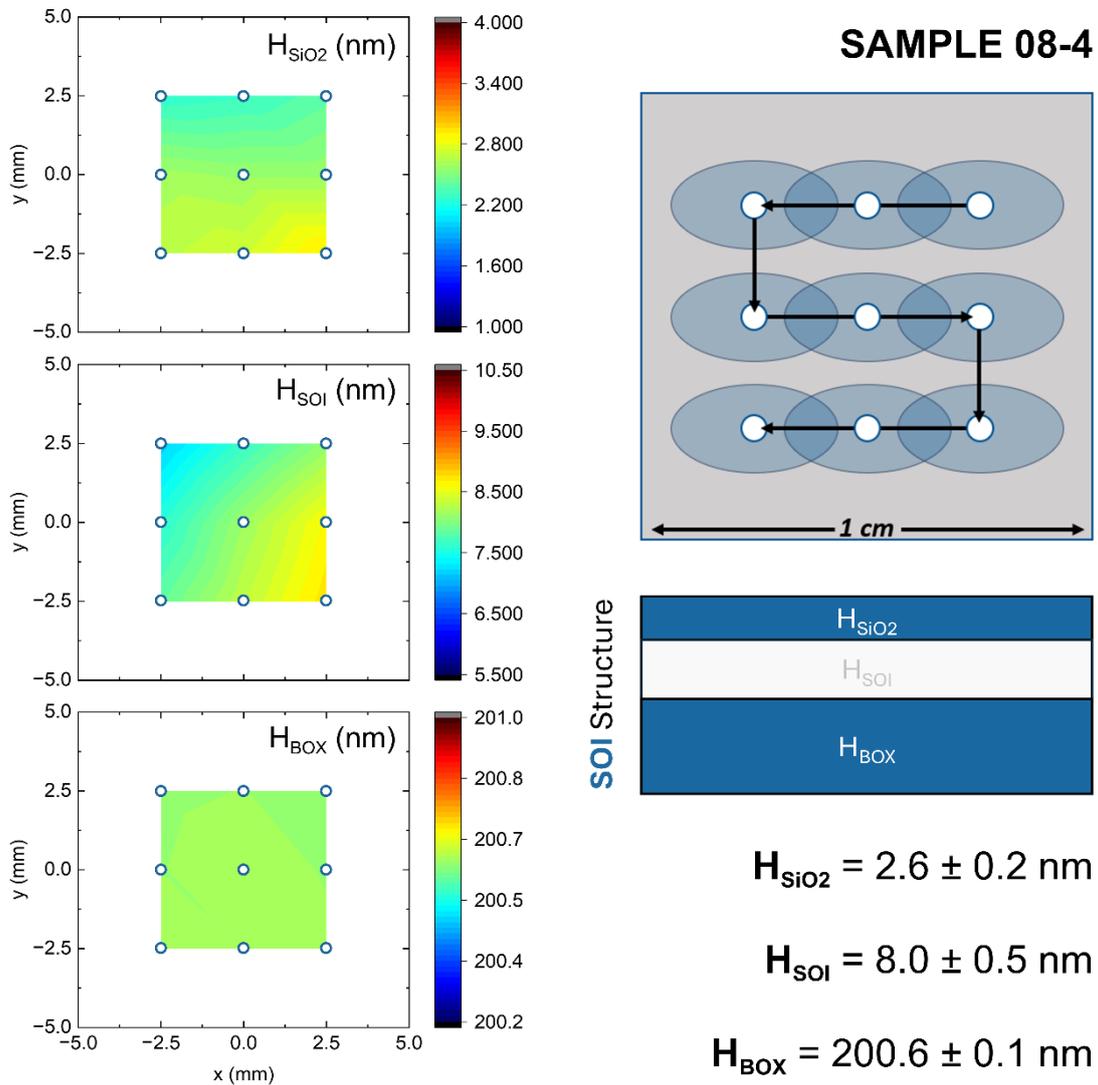

*Figure S1.* Spectroscopic ellipsometry (SE) measurements scheme of sample named 08-4, chosen as an example. Each SOI sample is measured at nine evenly spaced points across the center of the sample at a fixed incidence angle of 75°. Experimental data are fitted using a three-layer SOI model comprising the buried oxide (BOX), the device layer, and the top $SiO_2$ (TOX), as presented in the SOI structure. Mean values and errors are computed as the average and standard deviation of the nine points collected.

3. *Figure S2: Conductivity fit of the samples with $H_{SOI}$ ~ 30 nm*

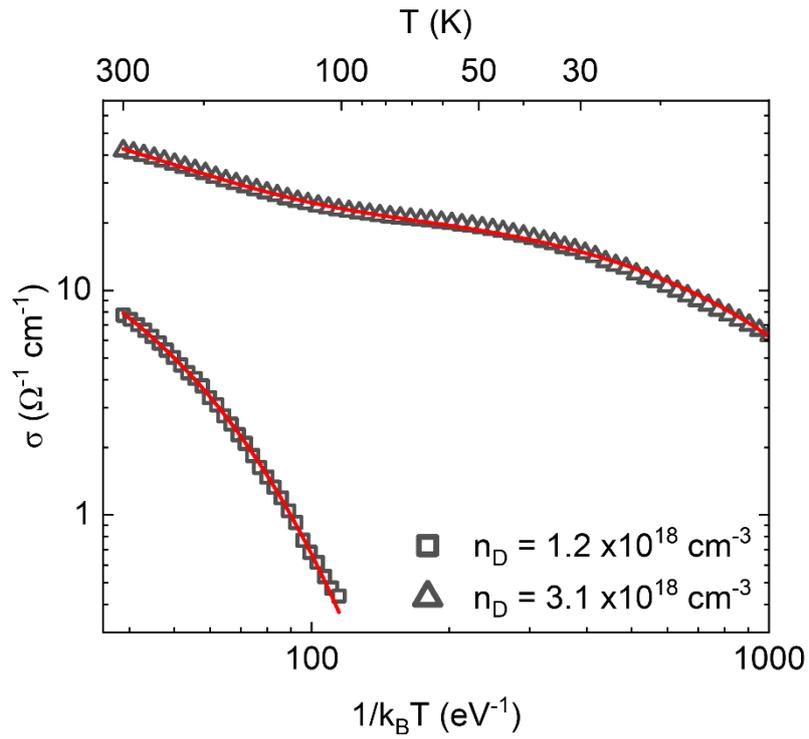

***Figure S2.*** *Conductivity ($\sigma$) as a function of the inverse of the temperature for the SOI samples with $H_{SOI}$ ~ 30 nm and $n_D = 1.2 \times 10^{18}$ cm$^{-3}$ and $n_D = 3.1 \times 10^{18}$ cm$^{-3}$. The red line corresponds to the fitting of the data using eq. 2.*

## 4. *Figure S3: Capacitance-Voltage (CV) measurements on Bulk Si*

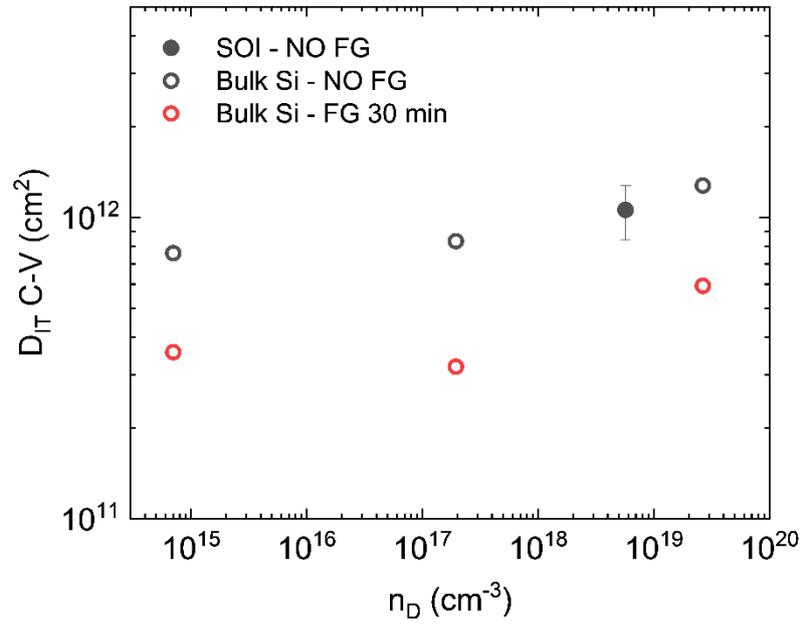

***Figure S3.*** *Interface state density ($D_{IT}$) obtained by CV measurements for RTO-SiO$_2$/Si TOX interfaces in bulk Si versus P concentration ($n_D$) before (black open circles) and after a 30 min forming gas anneal (red open circles). Values are compared to the average $D_{IT}$ extracted for SOI samples (black dot).*